\documentclass[twocolumn,showpacs,prb,aps,superscriptaddress]{revtex4}
\usepackage[latin1]{inputenc}
\usepackage{bm}
\usepackage{multirow}
\usepackage{amssymb}
\usepackage{amsbsy}
\usepackage{amsmath}
\usepackage{stmaryrd}
\usepackage{graphicx}
\usepackage{epsfig}

\begin{document}

\title{Density Dynamics in Translationally Invariant Spin-$1/2$ Chains at High Temperatures:
a Current Auto-Correlation Approach to Finite Time- and Length-Scales}

\author{Robin Steinigeweg}%
\email{rsteinig@uos.de}%
\affiliation{Fachbereich Physik, Universit\"at Osnabr\"uck, Barbarastrasse 7, %
             D-49069 Osnabr\"uck, Germany}%

\author{Jochen Gemmer}%
\affiliation{Fachbereich Physik, Universit\"at Osnabr\"uck, Barbarastrasse 7, %
             D-49069 Osnabr\"uck, Germany}%

\date{\today}

\begin{abstract}
We investigate transport in several translationally invariant
spin-$1/2$ chains in the limit of high temperatures. We
concretely consider spin transport in the anisotropic Heisenberg
chain, the pure Heisenberg chain within an alternating field, and
energy transport in an Ising chain which is exposed to a tilted
field. Our approach is essentially based on a connection between
the evolution of the variance of an inhomogeneous non-equilibrium
density and the current auto-correlation function at finite times.
Although this relationship is not restricted to the case of
diffusive transport, it allows to extract a quantitative value for
the diffusion constant in that case. By means of numerically exact
diagonalization we indeed observe diffusive behavior in the
considered spin chains for a range of model parameters and confirm
the diffusion coefficients which were obtained for these systems
from non-equilibrium bath scenarios.
\end{abstract}





\maketitle

%
%

\section{Introduction}
\label{introduction}

Although transport in low-dimensional quantum systems has
intensively been investigated theoretically in the past years,
there still is an ongoing interest in understanding the transport
phenomena in such systems, including their temperature and length
scale dependence \cite{zotos1999, narozhny1998, michel2003,
heidrichmeisner2003, rabson2004, heidrichmeisner2007, zotos2003,
benz2005, steinigeweg2006, santos2008, fabricius1998, sirker2009,
prosen2009, langer2009, prelovsek2004, michel2008, brandt1986,
roldan1986, sakai2003, saito2003, mejiamonasterio2005, mejiamonasterio2007,
karahalios2009, saito2000, gemmer2006, steinigeweg2007, wu2008,
rosch2000, steinigeweg2009-1, castella1995, steinigeweg2009-3}.
Those works have often addressed a qualitative classification of the
occurring transport types into ballistic or normal diffusive behavior
and, in particular cases, the crucial mechanisms which are responsible
for the emergence of diffusion have been studied. In this context the
role of non-integrability and quantum chaos is frequently discussed as
an at least necessary condition \cite{castella1995, rosch2000,
mejiamonasterio2005, steinigeweg2006, santos2008}. Significant
theoretical attention has been devoted to spin-$1/2$ chains
\cite{zotos1999, narozhny1998, michel2003, heidrichmeisner2003,
rabson2004, heidrichmeisner2007, zotos2003, benz2005,
steinigeweg2006, santos2008, fabricius1998, sirker2009, prosen2009,
langer2009, prelovsek2004, michel2008, brandt1986, roldan1986, sakai2003,
saito2003, mejiamonasterio2005, mejiamonasterio2007, karahalios2009},
e.g.~to the prominent anisotropic Heisenberg chain (XXZ model)
\cite{zotos1999, narozhny1998, michel2003, heidrichmeisner2003,
rabson2004, heidrichmeisner2007, zotos2003, benz2005, steinigeweg2006,
santos2008, fabricius1998, sirker2009, prosen2009, langer2009,
prelovsek2004, michel2008, brandt1986, roldan1986, sakai2003}. Most
controversial appears the question whether or not the (finite temperature)
transport in the pure Heisenberg chain is ballistic \cite{zotos1999,
narozhny1998, michel2003, heidrichmeisner2003, rabson2004,
heidrichmeisner2007, zotos2003, benz2005, steinigeweg2006, santos2008,
fabricius1998, sirker2009}.

Even though there certainly is a large variety of different
methods for the investigation of transport in quantum systems, we
concentrate here on two of the main approaches in detail which are also
most relevant in the context of the present work. The first approach
may be classified as a direct one, since it is rather close to an
experimental measurement setup: Within the theory of open quantum
systems \cite{breuer2007} the model of interest is coupled
locally, e.g.~at both ends of a spin chain, to reservoirs of the
transported quantity, e.g.~at different temperature or chemical
potential \cite{saito2000, saito2003, michel2003,
mejiamonasterio2005, mejiamonasterio2007, wichterich2007,
michel2008, prosen2009, steinigeweg2009-3}. Due to the coupling to
reservoirs, a stationary non-equilibrium state eventually results
for which all relevant expectation values such as the current and
the spatial density profile of the transport quantity can be
evaluated. Here, a vanishing profile corresponds to ballistic
behavior, whereas normal diffusive dynamics is associated with a
strictly linear profile, according to Fourier's law, respectively
Fick's law, see, e.g.~Ref.~\onlinecite{prosen2009}. In the latter
case a finite conductivity is simply given by the ratio of the
current and the spatial density gradient of the transported
quantity. This conductivity can be also understood in terms of a
diffusion coefficient, since transport is driven by a gradient
within the considered model and not by an external force
\cite{steinigeweg2009-2}. One may therefore directly compare with
the diffusion constant of a corresponding closed scenario, where
transport is not induced by the coupling to baths but by an
initially inhomogeneous non-equilibrium density
\cite{steinigeweg2009-2, steinigeweg2009-3}.

In order to simulate the influence of the baths the Liouville-von
Neumann equation for the coherent evolution of the density matrix
is routinely extended by an incoherent damping term, see,
e.g.~Ref.~\onlinecite{wichterich2007}. The derivation of such a
dynamical equation from a microscopic bath model is highly
non-trivial and involves a combination of various subtle
approximation schemes, e.g.~improper approximations may eventually
lead to a mathematically correct but physically irrelevant quantum
master equation ({\bf QME})\cite{wichterich2007}. One often
intends to derive a proper {\bf QME} of the Lindblad form
\cite{lindblad1976}, because its special structure allows to apply
the numerically efficient method of stochastic unraveling
\cite{dalibard1992, molmer1996}, e.g.~spin-$1/2$ chains with about
$16$ sites become numerically tractable \cite{michel2008}.
However, a recently suggested matrix product operator Ansatz has
been shown to significantly increase this number of available
sites up to several dozens \cite{prosen2009}, i.e.~finite size
effects are drastically reduced and the extracted conductivity may
be interpreted as a pure bulk property of the model, if it does
not depend crucially on the concrete form and strength of the bath
coupling, of course \cite{steinigeweg2009-3}.

Another approach for the investigation of transport in quantum
systems is the Green-Kubo formula ({\bf KF}) which was originally
derived for electrical conductance by the use of linear response
theory \cite{kubo1991, mahan2000}. In that case the
(frequency-dependent) electrical field is an external force which
perturbs the system and the resulting current of charge through
the system is the response to this external perturbation. The {\bf
KF} as such gives the (frequency-dependent) conductivity as the
linear response coefficient in terms of a current auto-correlation
function. The same approach is also used in the context of
gradient-driven transport phenomena, e.g.~for transport of energy
or heat. In that case the current is driven by a much more
complicated mechanism \cite{luttinger1964, zwanzig1965,
bonetto2000, lepri2003, garrido2001, gemmer2006, wu2008},
especially since it can not be treated as a perturbation to the
system. Nevertheless, simply by the replacement of the electrical
current, e.g.~by the energy current \cite{luttinger1964}, the {\bf
KF} is used for gradient-driven transport, too. But a rigorous
justification of this replacement remains still a conceptual
problem, see Ref.~\onlinecite{gemmer2006} or the overview paper
\onlinecite{zwanzig1965} (and the comprehensive literature which
is cited therein).

However, the {\bf KF} is nowadays a standard method for the
investigation of transport in spin chains \cite{narozhny1998,
zotos1999, zotos2003, heidrichmeisner2003, saito2003, rabson2004,
prelovsek2004, benz2005, heidrichmeisner2007, karahalios2009}, not
least due to its direct computability, once a finite piece of the
considered system has been exactly diagonalized. As far as numerics is
concerned, exact diagonalization is restricted to spin-1/2 chains
with at most 24 sites. But another difficulty arises, if the {\bf
KF} is evaluated on the basis of a finite piece: One often
distinguishes between ballistic and normal diffusive transport by
the notion of the Drude weight, essentially the conductivity's
singular contribution at zero-frequency, see the review
\onlinecite{zotos2003}, for example. Whenever it is finite, the
long-time behavior is expected to be ballistic, and whenever it
vanishes, the dc-conductivity as the zero-frequency limit of the
conductivity's regular non-singular part determines the long-time
behavior. Since for any finite system the conductivity exclusively
consists of delta peaks at different frequencies, there apparently
are only singular contributions and it therefore is a difficult
question how to extract and extrapolate the dc-conductivity from a
finite system \cite{gemmer2006}. Moreover, the quantitative
comparison of the resulting dc-conductivity (due to a possibly
hypothetic external force) with a diffusion constant (from
non-equilibrium bath scenarios) is hardly possible without the
detailed knowledge about an Einstein relation between both quantities.

In the present paper we do not intend to further discuss the above
mentioned conceptual and methodological problems which may come
along with the {\bf KF}. Instead we will present a different but
in a sense related approach which particularly is not concerned
with the most of those problems. To this end we will firstly
introduce a ``typical'' inhomogeneous non-equilibrium density in
Sec.~\ref{diffusion} and then connect the evolution of the
variance of this density to a current auto-correlation function at
finite times \cite{steinigeweg2009-2}. Remarkably, in the special
limit of infinitely long times this connection will be shown to
yield a generalized Einstein relation which relates the diffusion
constant to the dc-conductivity, i.e.~as evaluated by the {\bf
KF}. Moreover, we will demonstrate that the great advantage of the
connection is given by its direct applicability at finite times
and for finite systems as well. In this context we will suggest
another concept for the analysis of data which is available from
current auto-correlations functions.

By the use of the suggested concept we will investigate in
Secs.~\ref{tilted}-\ref{alternating} transport in several
translationally invariant spin-1/2 chains in the special limit of
high temperatures. We will concretely consider spin transport in
the anisotropic Heisenberg chain, the pure Heisenberg chain within
an alternating field, and energy transport in an Ising chain
which is exposed to a tilted field. By means of numerically exact
diagonalization we indeed observe strong indications for
diffusive behavior in the considered spin chains for a range of
model parameters and, what is more, we are able to quantitatively
confirm the diffusion constants which were found for these systems
from non-equilibrium bath scenarios in
Refs.~\onlinecite{mejiamonasterio2005, mejiamonasterio2007,
michel2008, prosen2009}. Finally, we will close in
Sec.~\ref{summary} with a summary and a conclusion.

%
%

\section{Connection between Variance and Current Auto-Correlation Function}
\label{diffusion}

In this Sec.~we are going to introduce our approach to
density-driven transport in translationally invariant quantum
systems. To this end the next Sec.~\ref{diffusion_a} firstly
presents the pertinent definitions and exclusively describes
the general theory which eventually yields the basic Eq.~(\ref{D})
for the time-dependent diffusion constant ${\cal D}(t)$. In the
following Sec.~\ref{diffusion_b} we then motivate to evaluate
${\cal D}(t)$ for finite times $t$ and particularly illustrate the
concept which will be used for the concrete spin chains in the
subsequent Secs. Thus, the reader which is not primarily interested
in the theoretical details may directly continue with
Sec.~\ref{diffusion_b}.

\subsection{Diffusion Constant}
\label{diffusion_a}

In the present paper we will investigate translationally invariant,
one-dimensional quantum spin systems which are described by a
respective Hamiltonian $\hat{H}$. In those quantum systems we will
consider an overall conserved transport quantity $\hat{X}$,
i.e.~$[ \hat{H}, \hat{X} ] = 0$. This transport quantity and the
Hamiltonian as well are decomposable into $N$ formally identical
addends $\hat{x}_\mu$, respectively $\hat{h}_\mu$ corresponding to
different positions, i.e.
\begin{equation}
\hat{X} = \sum_{\mu = 1}^N \hat{x}_\mu \, , \quad \hat{H} =
\sum_{\mu = 1}^N \hat{h}_\mu \, .
\end{equation}
Thus, $\hat{x}_\mu$ is a local density of the transported
quantity $\hat{X}$. Note that the $\hat{x}_\mu$ may be defined on
the positions of the $\hat{h}_\mu$, in between, or both. The above
decomposition is further done in such a way that Heisenberg's
equation of motion for the local densities $\hat{x}_\mu$ reads
\begin{equation}
\frac{\text{d}}{\text{d}t} \, \hat{x}_\mu = \imath \, [ \hat{H},
\hat{x}_\mu ] = \underbrace{\imath \, [ \hat{h}_{\mu^-},
\hat{x}_\mu ]}_{\equiv \hat{j}_{(\mu-1)}} + \underbrace{\imath \,
[ \hat{h}_{\mu^+}, \hat{x}_\mu ]}_{\equiv - \hat{j}_\mu} \, ,
\label{continuity}
\end{equation}
where $\hat{h}_{\mu^-}$ and $\hat{h}_{\mu^+}$ represent those
local addends of the Hamiltonian $\hat{H}$ which are located
directly on the l.h.s., respectively r.h.s.~of $\hat{x}_\mu$. This
apparently implies a kind of locality. However, such a description
can always be at least approximately enforced, if only
interactions are reasonably short-ranged. For all quantum systems
in the following Secs.~\ref{tilted}-\ref{alternating} the
description will be even exact, because interactions between
nearest-neighbors are taken into account solely. As routinely
done, the comparison with a continuity equation suggests the
definition of a local current $\hat{j}_\mu$ according to the
scheme in Eq.~(\ref{continuity}), see
Ref.~\onlinecite{gemmer2006}, for example. This definition is
consistent, if
\begin{equation}
\underbrace{\imath \, [ \hat{h}_{\mu^+}, \hat{x}_\mu ]}_{
-\hat{j}_\mu} + \underbrace{\imath \, [ \hat{h}_{(\mu+1)^-},
\hat{x}_{(\mu+1)} ]}_{\hat{j}_\mu} = 0 \, ,
\end{equation}
where the latter holds, if $\hat{X}$ is globally preserved. The
total current $\hat{J}$ is given by
\begin{equation}
\hat{J} = \sum_{\mu = 1}^N \hat{j}_\mu \, .
\end{equation}

Once the above decomposition has been established, we can define a
certain class of initial states $\rho(0)$ which corresponds to
some inhomogeneous, non-equilibrium density. To those ends let
\begin{equation}
\hat{d}_\mu \equiv \hat{x}_\mu - \langle \hat{x}_\mu \rangle
\end{equation}
denote the deviation of the local densities $\hat{x}_\mu$ from
their equilibrium average $\langle \hat{x}_\mu \rangle = \text{Tr}
\{ \hat{x}_\mu \rho_\text{eq} \}$, where $\rho_\text{eq}$ is any
stationary equilibrium state, i.e.~$[ \hat{H}, \rho_\text{eq}] =
0$. Then the initial state $\rho(0)$ reads
\begin{equation}
\rho(0) \equiv \rho_\text{eq} + \sum_{\mu = 1}^N
\frac{\delta_\mu}{\epsilon^2} \, \rho_\text{eq}^\frac{1}{2} \,
\hat{d}_\mu \, \rho_\text{eq}^\frac{1}{2} \label{initialstate}
\end{equation}
with some realization for the numbers $\delta_\mu$. The factor
$\epsilon^2$ is concretely given by
\begin{equation}
\epsilon^2 \equiv \frac{1}{N} ( \langle \hat{X}^2 \rangle - \langle
\hat{X} \rangle^2) \label{epsilon2}
\end{equation}
and therefore quantifies the equilibrium fluctuations of the
transported quantity $\hat{X}$.

For the special initial state $\rho(0)$ we now consider
the actual expectation values
\begin{equation}
d_\mu(t) \equiv \text{Tr} \{ \hat{d}_\mu(t) \rho(0) \} \, .
\end{equation}
It follows that
\begin{equation}
\sum_{\mu = 1}^N d_\mu(t) = \sum_{\mu = 1}^N \delta_\mu \equiv \delta \, ,
\end{equation}
i.e.~the sum $\delta$ of the numbers $\delta_\mu$ in
Eq.~(\ref{initialstate}) determines the sum of the actual
expectation values $d_\mu(t)$.

Of particular interest is the spatial variance $W^2(t)$ of the
$d_\mu(t)$. It is given by
\begin{equation}
W^2(t) \equiv \sum_{\mu = 1}^N \frac{d_\mu(t)}{\delta} \, \mu^2 -
\left [ \sum_{\mu = 1}^N \frac{d_\mu(t)}{\delta} \, \mu \right ]^2
\, . \label{definitionW2}
\end{equation}
If the dynamics of the $d_\mu(t)$ was indeed generated by a
discrete diffusion equation of the form
\begin{equation}
\frac{\text{d}}{\text{d} t} \, {d}_\mu(t) = {\cal D}(t) \left [ \,
d_{\mu-1}(t) - 2 \, d_\mu(t) + d_{\mu+1}(t) \right ] \, ,
\label{diffusionequation}
\end{equation}
then the evolution of this variance would read
\begin{equation}
\frac{\text{d}}{\text{d} t} \, W^2(t) = 2 \, {\cal D}(t) \, ,
\label{W2}
\end{equation}
as long as the $d_\mu(t)$ vanish at the ends of a chain (open
boundary conditions) or are reasonably concentrated at a sector of
a ring (closed boundary conditions). Even though
Eq.~(\ref{diffusionequation}) implies Eq.~(\ref{W2}), the inverse
direction generally is not true, of course.

However, in Ref.~\onlinecite{steinigeweg2009-1} a connection of the
form (\ref{W2}) has recently been found directly from
Eq.~(\ref{definitionW2}), i.e.~by firstly applying Heisenberg's
equation of motion to Eq.~(\ref{definitionW2}) and subsequently
manipulating the resulting equations. The time-dependent diffusion
constant ${\cal D}(t)$ reads
\begin{equation}
{\cal D}(t) = \frac{1}{N \, \epsilon^2} \int_0^t \text{d}t' \,
C(t') \label{D}
\end{equation}
and is essentially given in terms of a time-integral over the
current auto-correlation function
\begin{equation}
C(t) \equiv \text{Tr} \{ \hat{J}(t) \, \rho_\text{eq}^\frac{1}{2} \,
\hat{J} \, \rho_\text{eq}^\frac{1}{2} \} \, , \label{C}
\end{equation}
i.e.~$C(t) = \langle \hat{J}(t) \, \hat{J} \rangle$ in the limit
of high temperatures ($T \rightarrow \infty$). This limit will be
considered throughout this work.

Strictly speaking, the above diffusion constant ${\cal D}(t)$
is restricted to the special initial state $\rho(0)$ in
Eq.~(\ref{initialstate}), of course. Nevertheless, due to its
concrete form, this initial state represents an ensemble average
w.r.t.~to typicality \cite{goldstein2006, popescu2006,
reimann2007} or, more precisely, the dynamical typicality of
quantum expectation values \cite{bartsch2009}. Thus, the
overwhelming majority of all possible initial states with the same
$d_\mu(0)$ as $\rho(0)$ is also expected to approximately yield
the $d_\mu(t)$ corresponding to $\rho(0)$, if the dimension of the
relevant Hilbert space is sufficiently large, see especially
Ref.~\onlinecite{bartsch2009}. The latter largeness is certainly
fulfilled for all practical purposes. Or, in other words, the
concrete curves for $d_\mu(t)$ and thus for $W^2(t)$ only slightly
``fluctuate'' around the curve of the ensemble average. In fact,
for the single-particle quantum system in
Refs.~\onlinecite{michel2005, gemmer2006, steinigeweg2007,
steinigeweg2009-1} it has been demonstrated that
Eqs.~(\ref{W2})-(\ref{C}) correctly describe the dynamics for all
pure initial states $| \psi(0) \rangle$ which are not created
explicitly in order to violate these equations, see
Ref.~\onlinecite{michel2005, gemmer2006}.

Eventually, let us relate the non-perturbative result for the
diffusion constant ${\cal D}(t)$ to the standard result of linear
response theory for the dc-conductivity $\sigma_\text{dc}$
\cite{kubo1991, mahan2000}. In the limit of high temperatures this
conductivity may be written as
\begin{equation}
\sigma_\text{dc} = \frac{\beta}{N} \, \int_0^\infty \text{d}t' \,
C(t') \, , \label{sigmadc}
\end{equation}
where $\beta$ is the inverse temperature. The comparison of
Eq.~(\ref{sigmadc}) with Eq.~(\ref{D}) yields
\begin{equation}
{\cal D}(t \rightarrow \infty) = \frac{\sigma_\text{dc}}{\beta \,
\epsilon^2} \, ,
\end{equation}
i.e.~it leads to an Einstein relation. It is well-known that the
integral in Eq.~(\ref{sigmadc}) will diverge, whenever
$C(\omega)$, the Fourier transform of $C(t)$, has a finite
contribution at $\omega = 0$, see Ref.~\onlinecite{zotos2003}, for
example. As routinely done, we may hence define a Drude weight
$D$, e.g.
\begin{equation}
D \equiv \lim_{t \rightarrow \infty} \frac{{\cal D}(t)}{t} =
\frac{C(\omega = 0)}{N \, \epsilon^2} \, .
\end{equation}

\subsection{Finite Time and Length Scales}
\label{diffusion_b}

If the total current is strictly preserved, i.e.~$[\hat{H},
\hat{J} ] = 0$, the diffusion constant ${\cal D}(t)$ is completely
governed by the Drude weight $D$, i.e.~${\cal D}(t) = D \; t$ for
all $t$. Consequently, transport is purely ballistic at each time,
respectively length scale. Even if the total current does not
represent a strictly conserved quantity, a non-zero Drude weight
directly implies that the diffusion constant ${\cal D}(t)$ is
still approximately given by the straight line $D \, t$ in the
limit $t \rightarrow \infty$. However, at some finite time or
length scale the diffusion constant ${\cal D}(t)$ may nevertheless
appear to be almost constant, i.e.~${\cal D}(t) \approx {\cal D}$,
as expected for diffusive behavior. Such a behavior requires that
the Drude weight $D$ is not relevantly large (on the corresponding
scale), the finiteness of $D$ by itself is not crucial in this
context. In fact, it is well-known that the classification of
transport types for a given quantum system generally is a concept
which crucially depends on the considered time or length scale,
see Refs.~\onlinecite{gemmer2006, steinigeweg2007,
steinigeweg2009-1}, for example.

The above line of reasoning becomes also relevant in those
situations where some model parameter $\lambda$ induces a
transition from a finite towards a zero Drude weight $D$,
e.g.~where $D$ vanishes above some critical value of $\lambda$,
say $\lambda_C$. Then at this critical value $\lambda_C$ a sharp
transition from ballistic towards non-ballistic transport is to be
expected at the infinite time scale. However, such a transition
does not necessarily appear at some finite time, respectively
length scale. Even if there was a transition, a sharp one would
require at the critical value $\lambda_C$ a sudden jump of the
Drude weight $D$ from a finite and relevantly large number (on the
corresponding scale) to zero.

We hence do not concentrate merely on Drude weights and their
finiteness, although those are also discussed, of course. Instead
of that we focus on the diffusion constant ${\cal D}(t)$, as
defined in Eq.~(\ref{D}), at finite times and for finite systems
as well. According to Eq.~(\ref{D}), we can evaluate the
underlying current auto-correlation function directly in the time
domain, i.e.~$C(t)$, instead of the frequency domain,
i.e.~$C(\omega)$. This way we are not concerned with the problems
which may arise due to the fact that $C(\omega)$ is a highly
non-smooth function for a finite system. We particularly exploit
that the ${D\cal}(t)$-curve does not change at sufficiently short
time scales any more, when the size of a system becomes large
enough. Thus, at those short time scales, interesting signatures
of the infinitely large system may already be extractable for a
system with an accessible size. In the following
Secs.~\ref{tilted}-\ref{alternating} this concept will be
demonstrated in full detail.

%
%

\section{Ising Chain within a Tilted Field}
\label{tilted}

\begin{figure}[htb]
\includegraphics[width=1.0\linewidth]{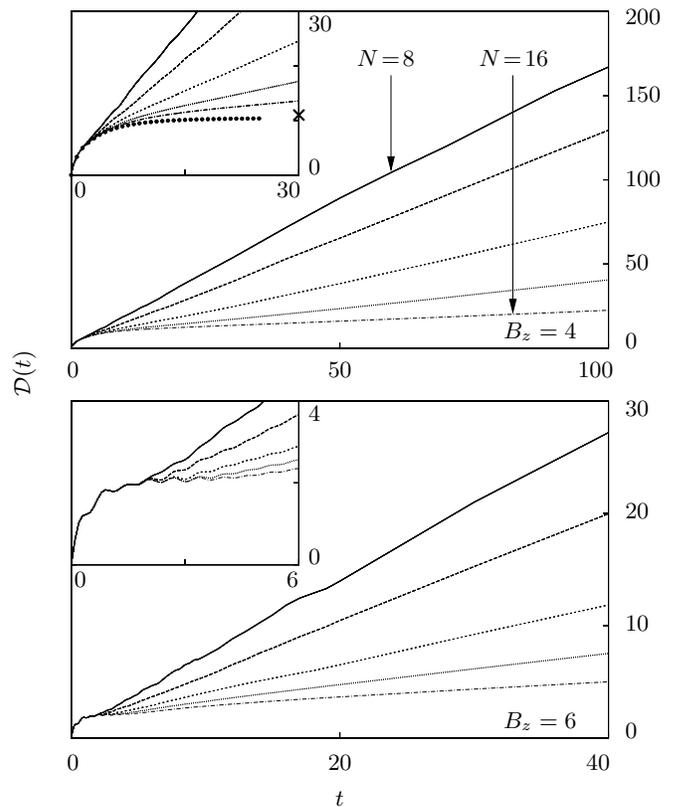}
\caption{The diffusion constant ${\cal D}(t)$, as given by
Eq.~(\ref{D}), for energy transport in the Ising chain within a
tilted field in the high temperature limit ($T = \infty$).
Insets zoom in ${\cal D}(t)$ at short $t$. Parameters: $J = -8$,
$B_x = 6.75$, $B_z = 4$ (top) as well as $B_z = 6$ (bottom). All
curves are evaluated numerically (exact diagonalization) for chain
lengths $N = 8$, $10$, $\ldots$, $16$ (arrows). The circles (top
inset) represent additional data for $N = 24$, extracted from
Ref.~\onlinecite{mejiamonasterio2005} and computed by the use of
approximative numerical integrators. The cross (top inset) indicates
the conductivity ${\cal D}_\text{bath}$ = 11 from
Ref.~\onlinecite{prosen2009} \cite{note1}, as obtained from a
non-equilibrium bath scenario for the same set of parameters.}
\label{D_tilted}
\end{figure}

\begin{figure}[htb]
\includegraphics[width=1.0\linewidth]{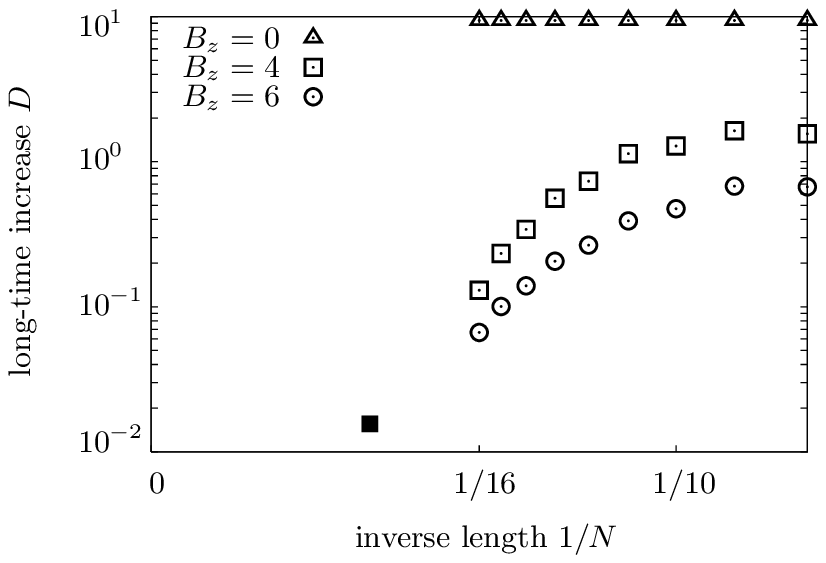}
\caption{The long-time increase $D$ of the diffusion constant
${\cal D}(t)$, the Drude weight, for energy transport in the Ising
chain within a tilted field at high temperatures ($T = \infty$).
Parameters: $J = -8$ and $B_x = 6.75$. All open symbols are
evaluated numerically (by the use of exact diagonalization) for
chain lengths $N \leq 16$ and for $z$-components $B_z \leq 6$. The
filled square is extracted from
Ref.~\onlinecite{mejiamonasterio2005} and represents an upper
bound for $N = 24$.} \label{Drude_tilted}
\end{figure}

In the present Section we will study a first example of a
translationally invariant spin system. This system is an Ising chain
which is exposed to a, say, tilted magnetic field $\bf B$.
Concretely, its Hamiltonian reads ($\hbar = 1$)
\cite{mejiamonasterio2005, mejiamonasterio2007, prosen2009}
\begin{equation}
\hat{H} = \sum_{\mu = 1}^N \hat{h}_\mu \, , \nonumber
\end{equation}
\begin{equation}
\hat{h}_\mu = \frac{J}{4} \, \hat{\sigma}_\mu^z
\hat{\sigma}_{\mu+1}^z + \frac{B_z}{4} \, (\hat{\sigma}_\mu^z +
\hat{\sigma}_{\mu+1}^z) + \frac{B_x}{4} \, (\hat{\sigma}_\mu^x +
\hat{\sigma}_{\mu+1}^x) \, , \label{H_tilted}
\end{equation}
where $B_z$ and $B_x$ denote the $z$-component, respectively
$x$-component of the total vector ${\bf B} = (B_x,0,B_z)$. Here,
one might think of some magnetic field which originally was in
line with the $z$-direction and has been rotated about the
$y$-axis with the angle $\alpha = \arctan(B_x/B_z)$.

In the Hamiltonian (\ref{H_tilted}) the operators
$\hat{\sigma}_\mu^i$ ($i = x, y, z$) are the standard Pauli
matrices (corresponding to site $\mu$); $J$ denotes the coupling
strength; and $N$ represents the total number of sites. Throughout
this work we use periodic boundary conditions, i.e. $N + 1 \equiv
1$.

Due to the presence of the tilted magnetic field, the only trivial
symmetries are translational invariance and mirror symmetry
\cite{mejiamonasterio2005} such that the whole Hilbert space is
decomposable into $2 \, N$ decoupled subspaces with similar
dimensions. For the present model the component of the total spin
$\hat{\bf S} = 1/2 \sum_{\mu = 1}^N \hat{\bf \sigma}_\mu$ in $\bf
B$-direction (and any other direction) is non-preserved. Therefore
magnetization (or spin) is not a suitable transport quantity here.

However, this model is appropriate to investigate the transport of
energy, i.e.~it allows to study the density dynamics of the local
quantities $\hat{h}_\mu$, see Eq.~(\ref{H_tilted}). Note that
these quantities contain contributions from both the Ising
interaction and the Zeeman energy. The respective energy current
$\hat{J}^E$ is given by \cite{mejiamonasterio2005,
mejiamonasterio2007, prosen2009}
\begin{equation}
\hat{J}^E = \sum_{\mu = 1}^N \hat{j}^E_\mu \, , \nonumber
\end{equation}
\begin{equation}
\hat{j}^E_\mu = \imath \, [\hat{h}_\mu, \hat{h}_{\mu+1}] = -
\frac{B_x \, J}{8} \, (\hat{\sigma}_\mu^z -
\hat{\sigma}_{\mu+2}^z) \, \hat{\sigma}_{\mu+1}^y \label{J_tilted}
\end{equation}
and the factor $\epsilon^2$, as defined in Eq.~(\ref{epsilon2}),
reads
\begin{equation}
\epsilon^2 = \frac{1}{16} \, (J^2 + 2 \, B_z^2 + 2\, B_x^2) \, .
\label{sigma2_tilted}
\end{equation}

If the component $B_z$ is identical to zero ($\alpha = 90^\circ$),
the energy current (\ref{J_tilted}) is strictly preserved.
Furthermore, the Hamiltonian (\ref{H_tilted}) is well-known to be
integrable. But it can become quantum chaotic for $B_z \neq 0$,
e.g.~for the special set of parameters $J = -8$, $B_z = 4$, $B_x =
6.75$ ($\alpha \approx 59^\circ$) \cite{mejiamonasterio2005,
mejiamonasterio2007, prosen2009}. Recently, for exactly this
parameter set, a strong evidence for diffusive behavior has been
found from non-equilibrium bath scenarios
\cite{mejiamonasterio2005, mejiamonasterio2007, prosen2009}.

Fig.~\ref{D_tilted} (top) shows the diffusion constant ${\cal
D}(t)$, as given by Eq.~(\ref{D}), for the above set of
parameters. We observe that ${\cal D}(t)$ increases within the
correlation time of the underlying current auto-correlation
function but, already on this time scale, essentially the Drude
weight governs the overall shape of the curve, at least for small
lengths $N \approx 8$. But when $N$ is increased, the Drude weight
$D$ rapidly becomes smaller, i.e $D$ decreases faster than a
power-law, see Fig.~\ref{Drude_tilted} (squares). In particular
there is no need to assume a finite and relevant value for $D$ in
the thermodynamic limit $N \rightarrow \infty$. Already for $N
\approx 16$ the Drude weight $D$ is such small that the tendency
of the diffusion constant ${\cal D}(t)$ to gradually develop
towards a horizontal line becomes visible. Nevertheless, only from
those lengths which are available from numerically exact
diagonalization a definite conclusion may still be vague,
particularly conclusions about the time after which ${\cal D}(t)$
possibly remains constant and about the constant value ${\cal D}$,
too.

Fortunately, additional data is available in literature: In
Ref.~\onlinecite{mejiamonasterio2005} the current auto-correlation
function has been evaluated also by the use of an approximative
numerical integrator. Based on this data the time evolution of the
diffusion constant ${\cal D}(t)$ can be extracted \cite{note2} for
the length $N = 24$, see Fig.~\ref{D_tilted} (top inset). And in
fact, for times which are larger than $t \approx 10$ we observe
that ${\cal D}(t)$ takes on a constant value ${\cal D} \approx
10.5$. The latter value further is in excellent agreement with
the conductivities ${\cal D}_\text{bath} \approx 10.3$ from
Ref.~\onlinecite{mejiamonasterio2005} and ${\cal D}_\text{bath} = 11$ from
Ref.~\onlinecite{prosen2009} \cite{note1}, as obtained therein for
the same set of parameters from non-equilibrium bath scenarios. (In
these works ${\cal D}_\text{bath}$ is denoted by $\kappa$.)

When the $z$-component is decreased from $B_z \approx 4$ down to
$0$, diffusive behavior eventually breaks down towards ballistic
transport, because the current becomes strictly preserved for $B_z
= 0$. Unfortunately, it is hardly possible to give a critical
value for this transition, simply due to the limited system sizes
which are accessible by the use of exact diagonalization. However,
such a value may be very unsharp for a continuous transition,
i.e.~if the Drude weight changes smoothly and does not jump
suddenly from zero (or a finite, irrelevantly small number) to a
finite, relevantly large number, cf.~Sec.~\ref{diffusion}.

Contrary, when the $z$-component is increased from $B_z \approx
4$, the indications of diffusive behavior become much more
pronounced for the accessible system sizes,
cf.~Fig.~\ref{D_tilted} (bottom). Respective diffusion constants
may therefore be suggested for this parameter regime.

%
%

\begin{figure}[htb]
\includegraphics[width=1.0\linewidth]{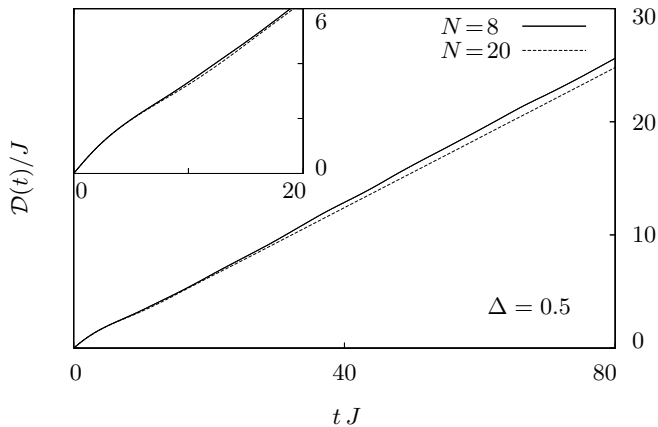}
\caption{The diffusion constant ${\cal D}(t)$, as given by
Eq.~(\ref{D}), for magnetization transport (spin transport) in the
anisotropic Heisenberg chain (XXZ model) in the high temperature
limit ($T = \infty$). Insets zoom in ${\cal D}(t)$ at short
$t$. The curves are evaluated numerically (by the use of exact
diagonalization) for the anisotropy parameter $\Delta = 0.5$ and
for chain lengths $N = 8$ and $20$.}
\label{D_anisotropic_0.5}
\end{figure}

\begin{figure}[htb]
\includegraphics[width=0.95\linewidth]{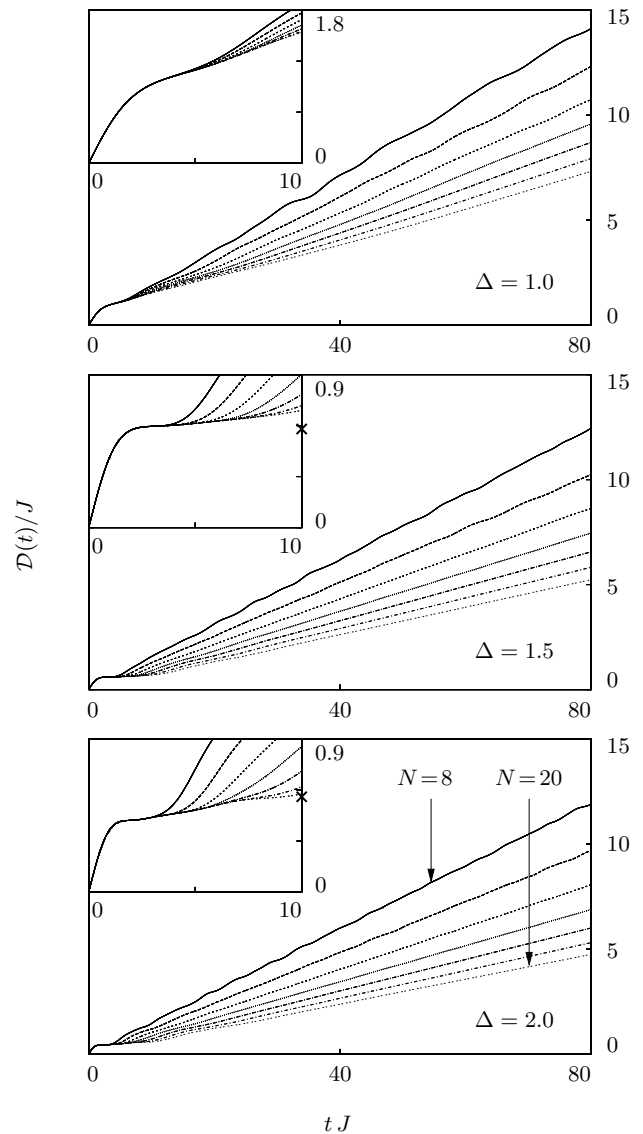}
\caption{The diffusion constant ${\cal D}(t)$, as given by
Eq.~(\ref{D}), for magnetization transport (spin transport) in the
anisotropic Heisenberg chain (XXZ model) in the high temperature
limit ($T = \infty$). Insets zoom in ${\cal D}(t)$ at short
$t$. The curves are evaluated numerically (by the use of exact
diagonalization) for chain lengths $N = 8$, $10$, $\ldots$, $20$
(arrows) and for anisotropy parameters $\Delta = 1.0$ (top), $1.5$
(middle), as well as $2.0$ (bottom). One cross (middle inset)
indicates the conductivity ${\cal D}_\text{bath}/J = 0.58$ in
Ref.~\onlinecite{prosen2009} \cite{note1}, as found therein from a
non-equilibrium bath scenario. Another cross (bottom inset)
represents the value $\sigma_\text{dc}/(\beta \epsilon^2 J) = 0.56$ with
the dc-conductivity $\sigma_\text{dc}/(\beta J) = 0.14$ according to
Ref.~\onlinecite{prelovsek2004}.} \label{D_anisotropic}
\end{figure}

\begin{figure}[htb]
\includegraphics[width=1.0\linewidth]{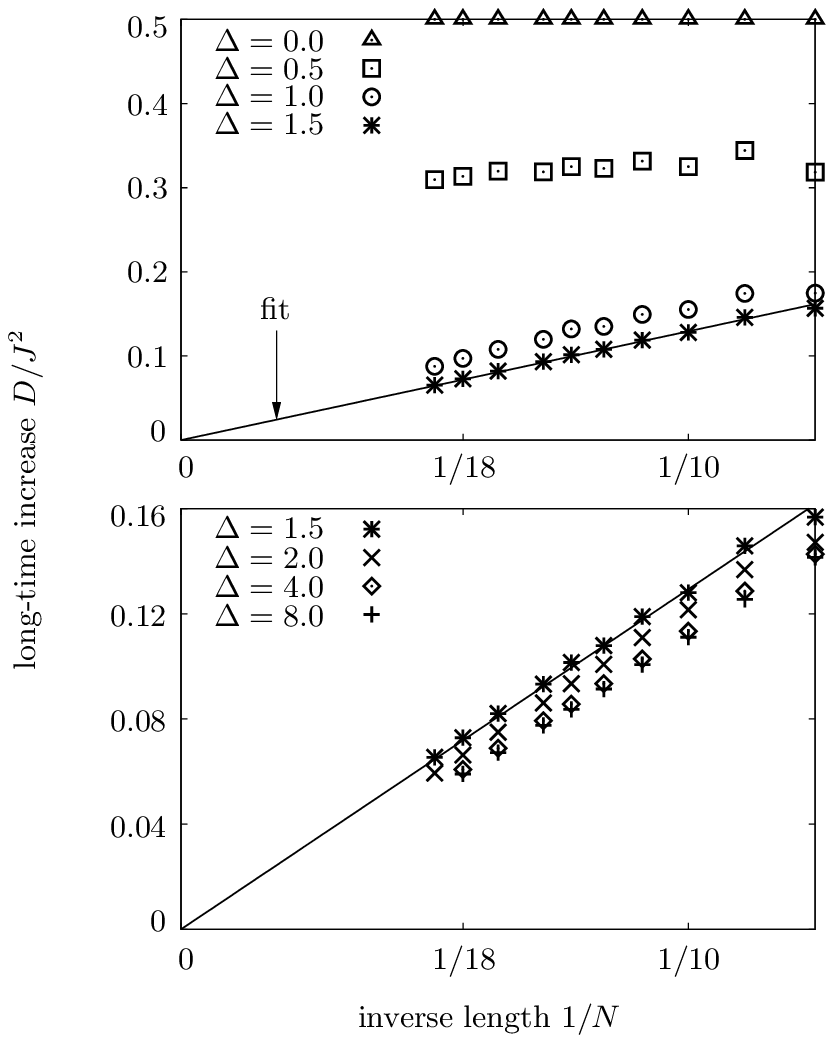}
\caption{The long-time increase $D$ of the diffusion constant
${\cal D}(t)$, the Drude weight, for magnetization transport (spin
transport) in the anisotropic Heisenberg chain (XXZ model) at high
temperatures ($T = \infty$). The symbols are evaluated numerically
(by the use of exact diagonalization) for chain lengths $N \leq
20$ and for anisotropy parameters $\Delta \leq 8$. The line
indicates a fit for the special case of $\Delta = 1.5$.}
\label{Drude_anisotropic}
\end{figure}

\section{Anisotropic Heisenberg Chain}
\label{anisotropic} In this Section we will investigate transport
in the anisotropic Heisenberg chain (or XXZ model) as another and
certainly more interesting example of a translationally invariant
spin system. It is described by a Hamiltonian of the form ($\hbar
= 1$) \cite{zotos2003, heidrichmeisner2007, michel2008,
prosen2009}
\begin{equation}
\hat{H} = \sum_{\mu = 1}^N \hat{h}_\mu \, , \nonumber
\end{equation}
\begin{equation}
\hat{h}_\mu = \frac{J}{4} (\hat{\sigma}_\mu^x
\hat{\sigma}_{\mu+1}^x + \hat{\sigma}_\mu^y \hat{\sigma}_{\mu+1}^y
+ \Delta \, \hat{\sigma}_\mu^z \hat{\sigma}_{\mu+1}^z)
\label{H_anisotropic}
\end{equation}
with the anisotropy parameter $\Delta$. Independent from the
concrete choice of $\Delta$, the Hamiltonian (\ref{H_anisotropic})
is integrable in terms of the Bethe Ansatz, see
Ref.~\onlinecite{zotos1999}, for example.

In the presence of an external (uniform) magnetic field $B$ one
may add to the Hamiltonian (\ref{H_anisotropic}) a Zeeman term of
the form $\hat{H}_B = B \, \hat{S}^z$. However, because the
following investigation (at infinite temperature) will not depend
on the concrete choice of $B$, we set $B = 0$ for simplicity.

As well-known, the Hamiltonian (\ref{H_anisotropic}) is invariant
under rotations about the z-axis, i.e.~it commutates with
$\hat{S}^z$. As a consequence $\hat{H}$ can be diagonalized within
decoupled subspaces with dimensions $N$ over $M + N/2$, where $M$
is the quantum number w.r.t.~$\hat{S}^z$. Moreover, due to the
translational invariance as well as the mirror symmetry of
(\ref{H_anisotropic}), the problem can be reduced further by a
factor $2 \, N$.

Since $\hat{S}^z$ represents a strictly conserved quantity, this
model allows to investigate the density dynamics of the local
quantities $\hat{\sigma}_\mu / 2$, i.e.~magnetization transport
(or spin transport). The respective magnetization current
$\hat{J}^S$ is given by \cite{zotos2003, heidrichmeisner2007,
 michel2008, prosen2009}
\begin{equation}
\hat{J}^S = \sum_{\mu = 1}^N \hat{j}^S_\mu \, , \nonumber
\end{equation}
\begin{equation}
\hat{j}^S_\mu = \imath \, [ \frac{1}{2} \hat{\sigma}_\mu^z,
\hat{h}_\mu ] = \frac{J}{4} (\hat{\sigma}_\mu^x
\hat{\sigma}_{\mu+1}^y -\hat{\sigma}_\mu^y \hat{\sigma}_{\mu+1}^x)
\, \label{J_anisotropic}
\end{equation}
and the factor $\epsilon^2$, as defined in (\ref{epsilon2}), takes
on the fixed value $1/4$.

Similarly, an energy current can be also introduced via the
relation $\hat{j}^E_\mu = \imath \, [ \hat{h}_\mu, \hat{h}_{\mu
+1} ]$. But, since its commutator with the Hamiltonian
(\ref{H_anisotropic}) vanishes exactly, this current is strictly
conserved such that energy transport is purely ballistic at each
time- and length-scale (and for all finite temperatures, see
Ref.~\onlinecite{sakai2003}).

Except for the special case of $\Delta = 0$ (XY model) the
magnetization current is non-preserved. On that account it could
be in principle possible to find diffusive behavior for non-zero
anisotropies. Nevertheless, in the literature there is strong
evidence that such a behavior is restricted to the regime $\Delta
> 1$ only \cite{narozhny1998, fabricius1998, zotos1999,
zotos2003, heidrichmeisner2003, rabson2004, benz2005,
heidrichmeisner2007, michel2008, langer2009, prosen2009}, most
promising appear those $\Delta$ which are close to $1.5$
\cite{michel2008, langer2009, prosen2009}. Those expectations are
also due to results on the Drude weight $D$ \cite{zotos1999,
benz2005, heidrichmeisner2003}, e.g.~Bethe Ansatz approaches
suggest that $D$ is finite for the regime $\Delta < 1$ and zero
for the regime $\Delta
> 1$ \cite{zotos1999, benz2005}. Still controversial is the
special case of $\Delta = 1$, where $D$ may be already zero
\cite{zotos1999} or not \cite{benz2005}.

For completeness, we also show numerical results for the Drude
weight in Fig.~\ref{Drude_anisotropic}, although respective data
can be found already in Ref.~\onlinecite{heidrichmeisner2003},
also obtained by the use of exact diagonalization and additionally
extrapolated to $N \rightarrow \infty$. This data can be
transferred directly to the present investigation, if the concrete
values for the Drude weight are multiplied by a factor
$1/(\epsilon^2 \pi) = 4/\pi$. Further results on Drude weights
from exact diagonalization can be found in,
e.g.~Refs.~\onlinecite{narozhny1998, rabson2004,
heidrichmeisner2007}.

\subsection{Anisotropies $\Delta < 1$}

Fig.~\ref{D_anisotropic_0.5} shows the diffusion constant ${\cal
D}(t)$, as defined in Eq.~(\ref{D}), for the anisotropy parameter
$\Delta = 0.5$. For $N = 8$ the Drude weight $D$ determines the
overall shape of the ${\cal D}(t)$-curve, i.e.~on all time scales
${\cal D}(t)$ is very close to the straight line $D \, t$. There
is only an insignificant deviation from this line at short times
scales below the correlation time of the underlying current
auto-correlation function, see Fig.~\ref{D_anisotropic_0.5}
(inset). For $N = 20$, i.e.~when the length of the chain is more
than doubled, the ${\cal D}(t)$-curve remains practically the
same. On that account it appears to be justified to assume a
similar curve in the thermodynamic limit $N \rightarrow \infty$.
This assumption is consistent with the extrapolation of the Drude
weight in Ref.~\onlinecite{heidrichmeisner2003}, see
Fig.~\ref{Drude_anisotropic} (squares), too. We may therefore
suggest purely ballistic behavior for the case $\Delta = 0.5$.

Note that Sirker et al.~have recently presented results
in Ref.~\onlinecite{sirker2009} which point towards a coexistence
of diffusive and ballistic dynamics, where the ballistic distribution
is small due to a small Drude weight. Even though we reproduce that
the Drude weight is large for $N \leq 20$ and does not depend
significantly on $N$, we can not exclude the possibility that the
Drude weight eventually becomes small for $N \gg 20$, of course.

\subsection{Anisotropies $\Delta > 1$}

In Fig.~\ref{D_anisotropic} (middle) we display the diffusion
constant ${\cal D}(t)$ for the anisotropy $\Delta = 1.5$. We
observe that ${\cal D}(t)$ increases at short time scales,
i.e.~within the correlation time of the underlying current
auto-correlation function $C(t)$, and then remains approximately
constant for an interval at intermediate time scales, i.e.~a
``plateau'' is formed at those times. Finally, a renewed increase
takes place on long time scales which is completely governed by
the Drude weight, or by the zero-frequency distribution of
$C(\omega)$. Note that the plateau can not be seen, if zero- and
finite-frequency parts are treated separately from each other. It
is a feature which arises from a combination of both
contributions.

The above plateau remarkably is already visible for $N = 8$.
Moreover, its ``height'' does not change with $N$, while its
``width'' seems to increase gradually, see
Fig.~\ref{D_anisotropic} (middle inset). The latter increase
particularly appears to be plausible, because the Drude weight is
commonly expected to vanish in the thermodynamic limit $N
\rightarrow \infty$, as already outlined above. We hence make the
educated guess that the plateau of ${\cal D}(t)$ will be continued
to, say, arbitrary long times, when only $N$ becomes sufficiently
large, e.g.~$N \rightarrow \infty$. Then the height ${\cal D}/J
\approx 0.60$ directly determines the concrete value of the
diffusion constant, of course. And indeed, the latter value for
$\cal D$ also is in excellent agreement with the conductivity
${\cal D}_\text{bath}/J = 0.58$ in Ref.~\onlinecite{prosen2009}
\cite{note1}, as found therein from a non-equilibrium bath scenario.
Although a respective figure is not shown for $\Delta = 1.6$, such
an agreement is additionally obtained with the results in
Ref.~\onlinecite{michel2008}, namely ${\cal D}/J \approx 0.55$ and
${\cal D}_\text{bath}/J = 0.585 \pm 0.020$. (In these works
${\cal D}_\text{bath}$ is denoted by $\kappa$.) Those quantitative
agreements support the correctness of our guess and the emergence
of diffusive transport appears to be verified for those $\Delta$
which are close to $1.5$.

Independent from the above arguments regarding the long-time
extrapolation for $N \rightarrow \infty$, we should mention that
for $N = 20$ the ${\cal D}(t)$-curves have already converged for
those times which are shorter than $t \approx 8 / J$, see
Fig.~\ref{D_anisotropic} (middle inset). Due to Eq.~(\ref{W2}),
this time corresponds to a length scale $W$ on the order of about
$3$ sites. Thus, at least for $W \leq 3$ the dynamics should be
known.

For larger anisotropy parameters $\Delta$ we find a similar
situation: Again there is a clearly developing plateau at
intermediate time scales, see Fig.~\ref{D_anisotropic} (bottom)
for the case $\Delta = 2.0$. For the accessible system sizes,
however, the final plateau height $\cal D$ does not seem to be
reached yet, i.e., $\cal D$ can not be read off with the same
accuracy, as done before for $\Delta \approx 1.5$. For comparison
we hence indicate in Fig.~\ref{D_anisotropic} (bottom inset) the
quantity $\sigma_\text{dc}/(\beta \epsilon^2 J) = 0.56$ with the
dc-conductivity $\sigma_\text{dc}/ (\beta J) = 0.14$ according to
Ref.~\onlinecite{prelovsek2004}, obtained therein from an analysis
on the basis of the standard Green-Kubo formula. The good
agreement with this result is noticeable.

\subsection{Anisotropy $\Delta = 1$}
As already mentioned above, it is still controversial, whether or
not the Drude weight is finite in the limit $N \rightarrow
\infty$, when $\Delta$ becomes exactly $1.0$. However, let us for
the moment assume that the Drude weight is indeed finite in that
limit. Then at the infinite time scale the diffusion constant
${\cal D}(t)$ is completely governed by this non-zero Drude weight
and increases linearly. At finite time scales, however, ${\cal
D}(t)$ may still appear to be almost constant and feature a
plateau like the one for the case $\Delta = 1.5$. In principle
such a plateau can be very wide, if only its height is large in
comparison with the Drude weight $D$ such that the finiteness of
$D$ is by itself not crucial in this context,
cf.~Sec.~\ref{diffusion}.

We hence show in Fig.~\ref{D_anisotropic} (top) the diffusion
constant ${\cal D}(t)$ for the special case $\Delta = 1$. The
initial short-time increase directly passes into the final
long-time increase, i.e in between a horizontal line is not
observable for the accessible system sizes. Nevertheless, we can
not exclude the possibility that a plateau with a height on the
order of ${\cal D}/J \approx 1.0$ will eventually develop, when
the system size is further increased, cf.~Fig.~\ref{D_anisotropic}
(top inset). We may thus compare this value for ${\cal D}$ with
the extrapolated Drude weight $D/J^2 \approx 0.025$ in
Ref.~\onlinecite{heidrichmeisner2003} \cite{note3}, for example.
(The latter extrapolated $D$ is about $1/4$ of the long-time slope
for $N = 20$, cf.~Figs.~\ref{D_anisotropic} and
\ref{Drude_anisotropic}.) By the use of the rough estimation $D \,
|t_2 - t_1| / {\cal D} \ll 1$ we directly obtain that an interval
with ${\cal D}(t) \approx {\cal D}$ can not be wider than some
time scale $|t_2 - t_1| \ll 40 / J$, respectively length scale
$|W_2 - W_1| \ll 9$. Thus, if the above extrapolation for the
Drude weight was indeed correct, diffusion would be restricted to
very few sites only. Contrary, if the Drude weight was zero (or a
finite, irrelevantly small number), diffusion could occur for a
possible arbitrary number of sites, of course.

%
%

\begin{figure}[htb]
\includegraphics[width=1.0\linewidth]{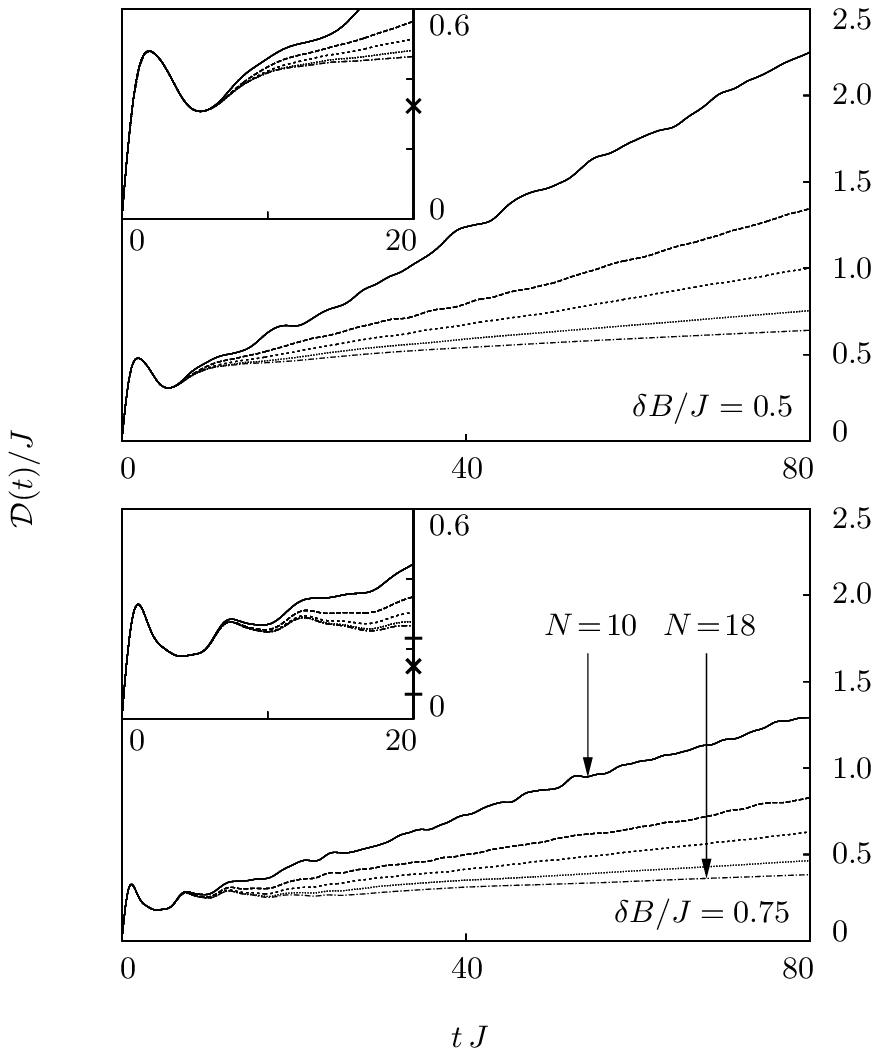}
\caption{The diffusion constant ${\cal D}(t)$, as given by
Eq.~(\ref{D}), for magnetization transport (spin transport) in the
Heisenberg chain within an alternating field in the high
temperature limit ($T = \infty$). Insets zoom in ${\cal D}(t)$
at short $t$. The curves are evaluated numerically (by the use of
exact diagonalization) for chain lengths $N = 10$, $12$, $\ldots$,
$18$ (arrows) and for field strengths $\delta B / J = 0.5$ (top) as
well as $0.75$ (bottom). The crosses (insets) indicate the conductivities
${\cal D}_\text{bath} /J = 0.323 \pm 0.010$ and $0.15 \pm 0.08$ in
Ref.~\onlinecite{michel2008}, as obtained from a non-equilibrium bath
scenario.}
\label{D_alternating}
\end{figure}

\begin{figure}[htb]
\includegraphics[width=1.0\linewidth]{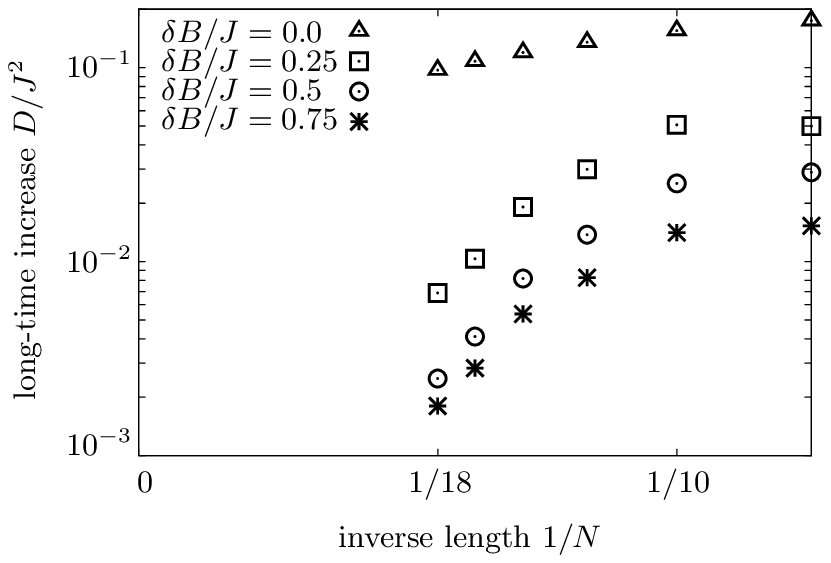}
\caption{The long-time increase $D$ of the diffusion constant
${\cal D}(t)$, the Drude weight, for magnetization transport (spin
transport) in the Heisenberg chain within an alternating field at
high temperatures ($T = \infty$). The symbols are evaluated
numerically (by the use of exact diagonalization) for chain
lengths $N \leq 18$ and for field strengths $\delta B / J \leq
0.75$.} \label{Drude_alternating}
\end{figure}

\section{Heisenberg Chain within An Alternating Field}
\label{alternating}

In this Section we are going to concentrate exclusively on the
pure Heisenberg chain, i.e.~$\Delta = 1$, and investigate another
modification of the Hamiltonian (\ref{H_anisotropic}) which may
lead to diffusive transport. Such a modification certainly is the
incorporation of disorder, i.e.~coupling constants or field
strengths which vary randomly from one site to another
\cite{avishai2002, santos2004, steinigeweg2006, santos2008, karahalios2009}.
But the investigation of disorder is numerically rather challenging
and conceptually more subtle due to the localization phenomenon,
even for a non-interacting single-particle system
\cite{anderson1958, steinigeweg2009-1}. Instead we will consider
another scenario without disorder, where the Heisenberg chain is
exposed to a strictly alternating field. This scenario is
concretely described by the Hamiltonian (\ref{H_anisotropic}) and
an additional Zeeman term of the form \cite{michel2008,
prosen2009}
\begin{equation}
\hat{H}_B = \sum_{\mu = 1}^N \frac{B + (-1)^\mu \, \delta B}{2} \,
\hat{\sigma}_\mu^z \, , \label{H_alternating}
\end{equation}
where $\delta B$ is the deviation from the mean $B$. As already
done before, we may set $B = 0$ for simplicity.

The presence of the Zeeman term (\ref{H_alternating}) does not
affect the commutation of $\hat{H}$ and $\hat{S}^z$, i.e.~the
Hamiltonian can be diagonalized within decoupled $M$-subspaces,
too. As long as $N$ is even, there also is translational
invariance (w.r.t two sites) such that the problem can be reduced
further by a factor $N/2$. (There is no mirror symmetry for even
$N$.)

Since $\hat{S}^z$ still represents a strictly conserved quantity,
the present model allows to investigate the transport of
magnetization w.r.t.~the field parameter $\delta B$. Note that the
respective current $\hat{J}^S$ and factor $\epsilon^2$ are
identical to those in Sec.~\ref{anisotropic}. When $\delta B$ is
increased from zero and becomes comparative with the coupling
strength $J$, the model undergoes a transition to quantum chaos
\cite{prosen2009}, i.e.~one may assume that the latter transition
already is a first pointer towards the onset of diffusion. Even
though it is entirely independent from those assumptions, we start
our investigation with $\delta B / J = 0.5$.

For $\delta B / J = 0.5$ the curve for the diffusion constant
${\cal D}(t)$ has changed from Fig.~\ref{D_anisotropic} (top) into
Fig.~\ref{D_alternating} (top). In particular the initial
short-time increase does not directly pass into the final
long-time increase any more. Instead there is an oscillation in
between such that, at the first view, the situation appears to be
much more complicated than a simple horizontal line. However, for
$N = 18$ the ${\cal D}(t)$-curve has already converged until the
end of this oscillation, see Fig.~\ref{D_alternating} (top inset).
Exactly at this position the curve seems to gradually develop a
plateau with a height on the order of ${\cal D} \approx 0.42 \,
J$. The plateau becomes visible, since the Drude weight rapidly
decreases with $N$, i.e.~it is about one order of magnitude
smaller for $N = 18$ than for $N = 8$,
cf.~Fig.~\ref{Drude_alternating} (circles). Note that the Drude
weight does not fulfill a simple $1/N$-dependence and decays
faster than a power-law. Thus, there is no need to suppose a
finite and relevant Drude weight in the limit $N \rightarrow
\infty$.

Remarkably, the above suggested ${\cal D}$ for $\delta B / J =
0.5$ agrees well with the conductivity ${\cal D}_\text{bath}/
\lambda = 0.323 \pm 0.010$ from Ref.~\onlinecite{michel2008}, as
found therein from a non-equilibrium bath scenario. The deviation
from this conductivity is on the order of $30 \%$ solely. However,
the small deviation may be explained as follows: In
Ref.~\onlinecite{michel2008} the conductivity is evaluated by the
use of the Lindblad equation, i.e., it is extracted from the
steady state of a finite chain which is at both ends weakly coupled
to baths at different temperatures. This weak bath-coupling is
essential for the validity of the Lindblad approach. But, for a
finite chain, a too weak bath-coupling may also yield a value
${\cal D}_\text{bath}$ which is lower than the correct value,
say, ${\cal D}$, see Ref.~\onlinecite{steinigeweg2009-3}. Since
in Ref.~\onlinecite{michel2008} the independence of the conductivity
from the bath-coupling strength is not analyzed in detail, we
suppose that the above deviation results from a slightly too weak
coupling of chain and reservoirs.

For larger field parameters $\delta B$ suggestions are less
reliable, since the initial oscillations of the ${\cal D}(t)$-curves
become much more pronounced, as already visible in Fig.~\ref{D_alternating}
(bottom). Even though there are oscillations, ${\cal D}(t)$ still
remains strictly positive, i.e.~those oscillations probably are no
pointer towards insulating behavior.

Note that definite conclusions can not be made for the parameter
regime $\delta B / J < 0.5$, since the curves for ${\cal D}(t)$
continuously change from Fig.~\ref{D_anisotropic} (top) into
Fig.~\ref{D_alternating} (top), i.e.~one basically is concerned
with the problem that the Drude weight governs the overall shape
of the curve, at least for accessible system sizes.

%
%

\section{Summary and Conclusion}
\label{summary}

In the present paper we have investigated transport in several
translationally invariant spin-$1/2$ chains in the special limit of
high temperatures. We have concretely considered spin transport in
the anisotropic Heisenberg chain, in the pure Heisenberg chain
within an alternating field, and energy transport in an Ising
chain which is exposed to a tilted field.

To this end we have firstly reviewed on a recently derived
connection between the evolution of the variance of some
``typical'' inhomogeneous non-equilibrium density and the current
auto-correlation function at finite times
\cite{steinigeweg2009-2}. In the limit of infinitely long times
this connection was shown to yield a generalized Einstein relation
which relates the diffusion constant (in the absence of any
external force) to the dc-conductivity (as the linear response
coefficient in the presence of an external force, i.e.~as
evaluated from the standard Green-Kubo formula \cite{kubo1991,
mahan2000}). However, we have additionally demonstrated that the
great advantage of the above connection is given by its direct
applicability at finite times and for finite systems, e.g.~at
short times interesting signatures of an infinitely large system
may be extractable for a system with an accessible size.

By means of numerically exact diagonalization we have indeed
observed strong indications for diffusive behavior in the
considered spin chains for a range of model parameters. Moreover,
the suggested diffusion constants have been found to be in
quantitative agreement with recent results on diffusion
coefficients which were obtained for the same spin chains from
numerically involved investigations of non-equilibrium bath
scenarios in Refs.~\onlinecite{mejiamonasterio2005,
mejiamonasterio2007, michel2008, prosen2009}.

Amongst all those and our findings at high temperatures the
emergence of diffusive transport of magnetization appears to be
verified in the anisotropic Heisenberg chain with the anisotropy
parameter $\Delta = 1.5$ \cite{michel2008, prosen2009}, despite
the integrability of the model. It is known that the onset of
quantum chaos is not a sufficient condition for diffusion
\cite{steinigeweg2006, santos2008} but the latter result suggests
also that non-integrability is not necessary at all.

Unfortunately, it is still an open question whether spin transport
in the pure Heisenberg chain is diffusive or ballistic. Simply by
the use of numerically exact diagonalization we were not able to
reach those system sizes which are required for any definite
conclusion on that question.

Therefore the next step certainly is the application of
approximate methods in order to obtain the current
auto-correlation function at finite times, either numerical ones,
e.g.~Suzuki-Trotter decompositions \cite{trotter1959, suzuki1990},
or analytical ones, e.g.~projection operator techniques
\cite{nakajima1958, zwanzig1960, breuer2007, steinigeweg2009-1} or
moment methods \cite{brandt1986, roldan1986}. Approximative methods
will also be in indispensable, when the investigation is extended
from one-dimensional spin chains to, e.g.~more-dimensional spin
lattices.

Apart from the above methodic details, a physically interesting
question is the dependence of transport on temperature, of course.
Since the connection between the evolution of the variance and the
current auto-correlation function is not restricted to infinite
temperature, further investigations may be done in this direction,
too. However, we expect that for low temperatures the convergence
of the diffusion constant at finite times is much slower, when the
system size is increased. Thus, for low temperatures conclusions
on the basis of a finite system may be less reliable.

%
%

\acknowledgments

We sincerely thank H.~Wichterich, M.~Michel and F.~Heidrich-Meisner for fruitful
discussions. We further gratefully acknowledge financial support by the Deutsche
Forschungsgemeinschaft.

%
%

\nocite{note1}%
\nocite{note2}%
\nocite{note3}%


\end{document}